\documentstyle[11pt]{article}

\begin{document}

\title{Black hole thermodynamics with generalized uncertainty principle}

\author{ Li Xiang\thanks{xiang.lee@163.com}~~and X. Q. Wen\\
 Center for Relativistic Astrophysics and High Energy Physics, \\Department of Physics, Nanchang University,\\ Nanchang,  330031,  Jiangxi province, P. R. China\\
 }
\date{}

\maketitle

\begin{abstract}
In the standard viewpoint, the temperature of a stationary black
hole is proportional to its surface gravity, $T_H=\hbar\kappa/2\pi$.
This is a semiclassical result and the quantum gravity effects are
not taken into consideration. This Letter explores a unified
expression for the black hole temperature in the sense of a
generalized uncertainty principle(GUP). Our discussion involves a
heuristic analysis of a particle which is absorbed  by the black
hole. Besides a class of static and spherically symmetric black
holes, an axially symmetric Kerr-Newman black hole is considered.
Different from the existing literature, we suggest that the black
hole's irreducible mass represent the characteristic size in the
absorption process. The information capacity of a remnant is also
discussed by Bousso's D-bound in de Sitter spacetime.


\end{abstract}
\newpage
\section{Introduction}
Heisenberg's uncertainty relation is  one of the fundamental
principles of quantum mechanics. This principle only meets the
quantum effects of matters, and it does not directly describe the
quantum fluctuations of spacetimes. However, many efforts have shown
that Heisenberg's principle may suffer a
modification\cite{yoneya1}-\cite{adler2}, in the context of quantum
gravity. Concretely, a generalized uncertainty principle(GUP) reads
\begin{eqnarray}\label{gup}
\Delta x \geq \frac{\hbar}{\Delta p}+\frac{\alpha}{\hbar} \Delta p,
\end{eqnarray}
where $\alpha\sim G$.   The second term on the r.h.s means a new
duality, which is firstly related to the spacetime uncertainty
principle\cite{yoneya1,yoneya2} and the scattering amplitude of high
energy string\cite{gross}. This term is also attributed  to gravity
in some gedanken experiments\cite{maggiore, garay, scard1, adler1}.
Different from Heisenberg's  principle, GUP restricts the shortest
distance that we can probe(i.e. $\Delta x\geq 2\sqrt{\alpha}\sim
l_p$). This agrees with the belief that Planck length is a
fundamental scale in quantum gravity.

Since uncertainty principle is of great importance to quantum
physics,  GUP has caused extensive interests and arguments. In
particular, GUP's effects on the thermodynamics of a Schwarzschild
black hole have been discussed by a heuristic method\cite{adler2}.
The crucial idea therein is that $\Delta x$ and $\Delta p$ are
identified as the black hole's size and temperature respectively. An
interesting result is that the black hole mass is not allowed to be
less than a scale of order Planck mass, which suggests a black hole
remnant. Although GUP's impacts on black hole thermodynamics have
been discussed in the literature\cite{adler2}-\cite{scard2}, a
universal expression is still absent.

In the semiclassical framework,  Hawking temperature of a stationary
black hole is proportional to the surface gravity, i.e.
\begin{eqnarray}\label{temp0}
T_H=\frac{\hbar\kappa}{2\pi},
\end{eqnarray}
where Planck constant reveals the quantum nature of black hole
radiation. In the Bekenstein's original work\cite{beken},
Heisenberg's uncertainty principle is crucial to the linear relation
between Hawking temperature and surface gravity.\footnote{This point
is also stressed in Ref.\cite{scard3}, where the linear relation
$T_H\sim \hbar\kappa$ can be obtained by another heuristic method
via Heisenberg's uncertainty principle.} In our opinion, GUP changes
the semiclassical framework to a certain context, and the
semiclassical black hole temperature (\ref{temp0}) should suffer a
modification. How the expression (\ref{temp0}) is corrected by GUP?
As an extension of Ref.\cite{adler2}, this research  explores a
revised temperature expression which is expected to be valid for
more general black holes. We  discuss a class of static and
spherically symmetric black holes, as well as a Kerr-Newman black
hole. The temperatures of these black holes have the same form. The
information capacity of a black hole remnant is discussed in de
Sitter spacetime, in terms of a Bousso's D-bound. We follow the
Bekenstein's original work\cite{beken}, and analyze a gedanken
experiment that a neutral particle just outside the horizon is
absorbed by the black hole. This Letter takes the units $G=c=k_B=1$.

\section{Black holes thermodynamics: a heuristic analysis}
\subsection{Brief review}
This subsection gives a brief review of the basis for the further
discussion. Let us start with the first law of black hole
mechanics\cite{beken, bardeen}
\begin{eqnarray}\label{firstlaw1}
dM=\frac{\kappa}{8\pi}dA+\sum_{i}Y_idy_i,
\end{eqnarray}
where the terms $\sum_i Y_idy_i$ represent the work done on the
black hole by an external agent. $y_i$ are the black hole's
variables such as electronic charge and angular momentum; $Y_i$  are
the generalized forces corresponding to the variables $y_i$,  e.g.
electrostatic potential and angular velocity. The above formula is a
result of classical general relativity. However, it has been endowed
with thermodynamic meaning since  Hawking radiation was discovered,
i.e.
\begin{eqnarray}
dM=TdS+\sum_{i}Y_idy_i.
\end{eqnarray}
Corresponding to the standard  temperature (\ref{temp0}), the black
hole entropy is expressed as $S_{BH}=(4\hbar)^{-1}A$, i.e. the
so-called Bekenstein-Hawking entropy. However, this simple relation
is a semiclassical result. In more general situations, the entropy
of a black hole is assumed to be a function of its area\cite{beken},
$S=S(A)$. Following from (\ref{firstlaw1}) and the definition of
thermodynamics, the temperature is expressed as
\begin{eqnarray}\label{temp1}
T=\left(\frac{\partial M}{\partial S}\right)_{y_i}=\frac{d A}{d
S}\times\left(\frac{\partial M}{\partial A}\right)_{y_i}=\frac{d
A}{d S}\times\frac{\kappa}{8\pi},
\end{eqnarray}
where the variables $y_i$ are fixed. The temperature expression is
determined by the relation between the entropy and  area. In order
to find the concrete form of $S(A)$, we consider a particle captured
by the black hole.  When the particle disappears, on one hand, its
information is lost to an observer outside the horizon; on the other
hand,  the smallest increase in the area of a Kerr-Newmann black
hole is given by\cite{beken}
  \begin{eqnarray}\label{minia0}
 \Delta A\sim b\mu,
 \end{eqnarray}
 where $b$ and $\mu$ are the particle's size and mass, respectively.
  Identifying the loss of
information with the increase of black hole entropy, we obtain
\begin{eqnarray}
\Delta S\simeq\frac{dS}{dA}\Delta A.\nonumber
\end{eqnarray}
 According to information theory, the loss of information is one
bit at least, i.e. $(\Delta S)_{min}=\ln 2$. The next step is to
work out the differential relation $dS/dA$ via (\ref{minia0}).  For
a classical particle(point-like object), $(\Delta A)_{min}=0$.
However, in quantum mechanics,
 a particle is described by a wave packet and a definite trajectory does not exist. The width of wave packet
  is defined as the standard deviation of $x$ distribution(i.e. the position
  uncertainty), which can be interpreted  as  the characteristic size of the particle($b\sim\Delta x$).
 Furthermore, the momentum uncertainty is not allowed to be greater than the
 mass ($\Delta p\leq\mu$), in the process of measuring the particle's position.
 Otherwise the relativistic effects lead to the creation of
a partner of the particle and make the measurement  meaningless.
Thus the expression (\ref{minia0}) is deduced to
\begin{eqnarray}\label{minia}
 \Delta A \sim b\mu\geq\Delta x\Delta p.
 \end{eqnarray}
The smallest increase in area cannot be arbitrarily small and it is
restricted by the uncertainty relation of quantum mechanics. In the
 Bekenstein's insightful work, Heisenberg principle is utilized to identify the particle's size
 with the Compton wavelength of itself,
  and then the
minimum increase in
 horizon area is given by $\Delta A\sim l_p^2$. This results in
\begin{eqnarray}
\frac{\Delta S}{\Delta A}=const,\nonumber
\end{eqnarray}
which means the linear relation between the black hole entropy and
the horizon area. GUP  will correct the Bekenstein's result.
Substituting (\ref{gup}) into (\ref{minia}), we have
\begin{eqnarray}\label{minia2}
 \Delta A \geq \gamma_1\hbar\left[1+\frac{\alpha}{\hbar^2}(\Delta
 p)^2\right],
 \end{eqnarray}
where $\gamma_1$ is a calibration factor. The minimum increase in
area, $(\Delta A)_{min}$, is determined by the  smallest uncertainty
of momentum. Following from (\ref{minia2}), $(\Delta A)_{min}$ would
be a constant if $\Delta p\rightarrow 0$. At a first glimpse, there
seems to be no correction to the Bekenstein's result. However,
$\Delta p\rightarrow 0$ means $\Delta x\rightarrow\infty$. For a
particle captured by black hole, $\Delta p$ is not allowed to be
arbitrarily small, since the particle is confined within a finite
region and $\Delta x$ is finite. $(\Delta A)_{min}$ is therefore no
longer a constant, which results in some corrections to the linear
relation between entropy and area. In the following subsections, a
static and spherically symmetric black hole as well as an axially
symmetric Kerr-Newman black hole are discussed respectively.
\subsection{A class of static and spherically black holes}
We consider a static and spherical black hole as follows
\begin{eqnarray}
ds^2=-F(r)dt^2+F^{-1}(r)dr^2+r^2(d\theta^2+\sin^2\theta
d\phi^2),\nonumber
\end{eqnarray}
where the horizon is located by $F(r_0)=0$.  The above line element
describes a class of static and spherically symmetric black holes,
such as Schwarzschild, Reissner-Nordstr\"{o}m and their partners in
(anti-)de Sitter spacetime. When a particle is captured by  black
hole, the position uncertainty  should not be greater than a
specific scale. This characteristic size, for a static and
spherically symmetric black hole, is identified with the twice
radius of horizon,\footnote{For example, see Ref.\cite{adler2}. In a
rotating case, the characteristic size is represented by the black
hole's irreducible mass. This point will be discussed in the next
subsection.} i.e.
\begin{eqnarray}\label{restrict1}
2r_0\geq\Delta x\geq \frac{\hbar}{\Delta p}+\frac{\alpha}{\hbar}
\Delta p,
\end{eqnarray}
which imposes a constraint on the momentum uncertainty as follows
\begin{eqnarray}\label{plimit}
\frac{\hbar}{\alpha}\left[r_0-\sqrt{r_0^2-\alpha}\right]\leq\Delta
p\leq \frac{\hbar}{\alpha}\left[r_0+\sqrt{r_0^2-\alpha}\right].
\end{eqnarray}
 So the product of $\Delta x$ and $\Delta p$  yields
\begin{eqnarray}\label{htype}
\Delta x\Delta p&\geq& \hbar\left[1+\frac{\alpha}{\hbar^2}(\Delta p)^2\right]\nonumber\\
&\geq&\frac{2\hbar}{\alpha}\left(r_0^2-r_0\sqrt{r_0^2-\alpha}\right)=\hbar^{\prime},
\end{eqnarray}
where the second inequality is obtained by taking the lower bound of
$\Delta p$. The above inequality can be rewritten as a
Heisenberg-type uncertainty principle, $\Delta x\Delta
p\geq\hbar^{\prime}$, where $\hbar^{\prime}$ may be regarded as an
effective Planck constant. Thus the increase in  area satisfies
\begin{eqnarray}\label{miniarea}
\Delta
A\geq\gamma_1\hbar^{\prime}=\frac{2\gamma_1\hbar}{\alpha}\left(r_0^2-r_0\sqrt{r_0^2-\alpha}\right).
\end{eqnarray}
 When the particle vanishes,
   the information of one bit is lost and the black hole acquires the increase in entropy
   $(\Delta S)_{min}=\ln 2$.
On the other hand, the minimum increase in the horizon area is given
by the lower bound of (\ref{miniarea}), which is denoted by $(\Delta
A)_{min}$. We obtain
\begin{eqnarray}\label{dads}
\frac{dA}{dS}\simeq \frac{(\Delta A)_{min}}{(\Delta
S)_{min}}=\frac{2\gamma_1\hbar}{\alpha\ln
2}(r_0^2-r_0\sqrt{r_0^2-\alpha}).
\end{eqnarray}
The black hole temperature (\ref{temp1}) is deduced to
\begin{eqnarray}
T\simeq\frac{\kappa}{8\pi}\cdot\frac{2\gamma_1\hbar}{\alpha\ln
2}(r_0^2-r_0\sqrt{r_0^2-\alpha}),\nonumber
\end{eqnarray}
which is not only proportional to the surface gravity but also
depends on the black hole size. It should reproduce the standard
result $T=\kappa/2\pi$, as $\alpha\rightarrow 0$. This requires that
the calibration factor yield $\gamma_1=4\ln 2$. Thus we obtain
\begin{eqnarray}\label{temp2}
T\simeq\frac{\hbar^{\prime}\kappa}{2\pi},
\end{eqnarray}
which is the expression for the temperature of a static and
spherically symmetric black hole.
 Comparing the standard formula (\ref{temp0}) with
the revised version (\ref{temp2}), we find that the latter can be
obtained from the former by substituting  $\hbar^{\prime}$ for the
Planck constant. It suggests that $\hbar^{\prime}$ play the role of
an effective Planck constant.

The expression (\ref{temp2}) can be understood by reexamining the
efficiency of a Geroch process. This gedanken experiment imagines a
machine operating between a black hole and a remote
reservoir.\footnote{For details, see Ref.\cite{beken}.}  In this
process, a box is filled with black body radiation from the
reservoir and lowered down to the black hole surface. After emitting
the radiation into the black hole, the box is moved away  from the
black hole. The over-all process converts heat into work with the
efficiency\cite{beken}
\begin{eqnarray}\label{effic}
\eta=1-\gamma_2\kappa\ell,
\end{eqnarray}
where $\ell$ is the size of the box, and $\gamma_2$ is a coefficient
factor to be determined. The smaller $\ell$ is, the greater $\eta$
is. In practical situations, it is reasonable that the box's size is
required to yield $\ell\leq 2r_0$. This is also necessary to emit
the total radiation into the black hole, otherwise the photons with
lower energy will not contribute to the Geroch process. On the other
hand, the box must have a nonzero size, and $\ell$ has a minimum
value which is related to the temperature of radiation, $T_R$. To
find the relation between the temperature and efficiency, we rewrite
(\ref{effic}) as
\begin{eqnarray}\label{effic2}
\eta=1-\gamma_2(\ell T_R)\frac{\kappa}{T_R}.
\end{eqnarray}
For a given reservoir, the maximum value of the efficiency is
determined by the smallest value for the production of $\ell$ and
$T_R$. As the characteristic energy of thermal photons, the
radiation temperature
 yields  $T_R>\epsilon$,  where $\epsilon$ is the photon's minimum energy which is given by the
lower bound of (\ref{plimit}). Thus we obtain
\begin{eqnarray}
\ell T_R>\epsilon\ell
&=&\frac{\hbar}{2\alpha}\left(\ell^2-\ell\sqrt{\ell^2-4\alpha}\right)\nonumber\\
&\geq&\frac{2\hbar}{\alpha}\left(r_0^2-r_0\sqrt{r_0^2-\alpha}\right)\nonumber\\
&=&\hbar^{\prime},\nonumber
\end{eqnarray}
where we have considered $\ell\leq 2r_0$. Thus the efficiency
(\ref{effic2}) yields
\begin{eqnarray}\label{effic3}
\eta<1-\gamma_2\frac{\hbar^{\prime}\kappa}{T_R}.
\end{eqnarray}
Comparing it with the efficiency of a heat engine operating between
two reservoirs,  we find that the expression $\hbar^{\prime}\kappa$
plays the role of the black hole temperature. This agrees with
(\ref{temp2}), up to a constant factor.

The black hole entropy can be expressed as
\begin{eqnarray}
S=\int\frac{dS}{dA}dA&\simeq&\int\frac{(\Delta S)_{min}}{(\Delta
A)_{min}}dA.\nonumber
\end{eqnarray}
Considering (\ref{dads}) and setting $\gamma_1=4\ln 2$, we obtain
\begin{eqnarray}\label{entropy}
S&\simeq&\frac{1}{4}\int\frac{dA}{\hbar^{\prime}}\nonumber\\
&=&\frac{\pi}{\hbar}\int\left(r_0+\sqrt{r_0^2-\alpha}\right)dr_0\nonumber\\
&=&\frac{\pi}{2\hbar}\left[r_0^2+r_0\sqrt{r_0^2-\alpha}-\alpha\ln(r_0+\sqrt{r_0^2-\alpha}
)\right].
\end{eqnarray}
When $r_0\gg\sqrt{\alpha}$, Bekenstein-Hawking entropy and the
log-type correction, as the first two leading terms in Taylor
series, are presented as
\begin{eqnarray}
S=(4\hbar)^{-1}(A-\alpha\pi\ln A+\cdots),\nonumber
\end{eqnarray}
where the log-type correction is similar to the existing results
that are derived from some concrete black holes by other
methods\cite{page}-\cite{maj3}.

 In the context of the GUP, the
  heat capacity is given by
\begin{eqnarray}\label{inheat}
C&=&T\frac{\partial S}{\partial
T}=\frac{\hbar^{\prime}\kappa}{2\pi}\cdot\frac{\partial S}{\partial
A}\cdot\frac{\partial
A}{\partial T}\nonumber\\
&=&\frac{1}{4}\left(\frac{\partial\hbar^{\prime}}{\partial
A}+\hbar^{\prime}\kappa^{-1}\frac{\partial\kappa}{\partial
A}\right)^{-1}.
\end{eqnarray}
 Direct calculation gives
\begin{eqnarray}
\frac{\partial\hbar^{\prime}}{\partial A}=\frac{1}{8\pi
r_0}\frac{\partial\hbar^{\prime}}{\partial r_0}=-\frac{\Delta
\hbar}{4f}.\nonumber
\end{eqnarray}
where $\Delta\hbar=\hbar^{\prime}-\hbar, ~f=f(r_0)=\pi
r_0\sqrt{r_0^2-\alpha}$.  The  heat capacity (\ref{inheat}) is
deduced to
\begin{eqnarray}\label{heat}
C=C_0f\left[\frac{\hbar^{\prime}}{\hbar}f-C_0\Delta\hbar\right]^{-1},
\end{eqnarray}
where
\begin{eqnarray}
C_0=T_H\frac{\partial S_{BH}}{\partial
T_H}=(4\hbar)^{-1}\kappa\frac{\partial A}{\partial\kappa},\nonumber
\end{eqnarray}
 is the standard heat capacity defined by the aid of
Hawking temperature (\ref{temp0}) and Bekenstein-Hawking
entropy.

Let us give a remark on the temperature expression (\ref{temp2}). In
Ref.\cite{adler2}, $\Delta x$ and $\Delta p$ are identified with the
black hole's radius and temperature respectively.
 However, this suggestion leads to a deduction that the temperature depends only on
  the black hole size. So the method of Ref.\cite{adler2} cannot be applied
to more cases with the exception of a Schwarzschild black hole. In
order to explain this weakness, let us observe a
Reissner-Nordstr\"{o}m black hole in de Sitter spacetime. Its
horizon radius $r_{0}$ is determined by
\begin{eqnarray}\label{rns1}
0=F(r_{0})=1-\frac{2M}{r_{0}}+\frac{Q^2}{r_{0}^2}-\frac{\Lambda}{3}r_{0}^2,
\end{eqnarray}
which has a very complex solution. However, we have no need  to know
the concrete form of $r_{0}$, because it is useless to our
discussion. Following from (\ref{rns1}), the mass is expressed as
\begin{eqnarray}\label{rns2}
M=\frac{1}{2}\left(r_{0}+\frac{Q^2}{r_{0}}-\frac{\Lambda}{3}r_{0}^3\right),
\end{eqnarray}
and the surface gravity is
\begin{eqnarray}\label{rns3}
\kappa=\frac{F^{\prime}(r_{0})}{2}=r_{0}^{-1}\left(\frac{M}{r_{0}}-\frac{Q^2}{r_{0}^2}-\frac{\Lambda}{3}r_{0}^2\right),
\end{eqnarray}
which is identified with the black hole temperature  in the
semiclassical framework. However, following from Ref.\cite{adler2},
the GUP (\ref{gup}) would give the temperature as follows
\begin{eqnarray}
T&\sim&
\frac{\hbar}{\alpha}\left[r_{0}-\sqrt{r_{0}^2-\alpha}\right]\nonumber\\
&\simeq&\hbar
r_{0}^{-1}\left(1+\frac{3\alpha}{4r_{0}^2}\right),\nonumber
\end{eqnarray}
which cannot reproduce the standard result( as $\alpha\rightarrow
0$). Therefore, (\ref{temp2}) is a nontrivial extension of
Ref.\cite{adler2}, since it is  suitable for more black holes and
can produce the standard expression.

 In the derivation of (\ref{temp2}), a hidden assumption is that the black holes  yield the laws (\ref{firstlaw1})
and (\ref{minia0}), including those in de Sitter spacetime. This is
an extension of Ref.\cite{beken}. There are some evidences for this
assumption. For example, following from  (\ref{rns2}) and
(\ref{rns3}), we obtain
\begin{eqnarray}\label{firstlaw2}
dM=\frac{\kappa}{8\pi}dA+\frac{Q}{r_{0}}dQ,
\end{eqnarray}
which is the first law of a Reissner-Nordstr\"{o}m black hole in de
Sitter spacetime. When the  black hole captures a neutral particle,
the first law becomes
\begin{eqnarray}\label{refirstlaw2}
dM=\frac{\kappa}{8\pi}dA.
\end{eqnarray}
On the other hand, based on a Bekenstein-type analysis, the smallest
increase in the black hole mass is given by\cite{zhaoz}
\begin{eqnarray}\label{dm}
\Delta M\sim b\mu\kappa,
\end{eqnarray}
where $b$ and $\mu$ are the particle's size and mass respectively.
 Considering (\ref{refirstlaw2}) and (\ref{dm}), the smallest
increase in horizon area is $\Delta A\sim b\mu$, which is just
(\ref{minia0}).

\subsection{Kerr-Newman black hole}
 In Boyer-Lindquist coordinates, a
Kerr-Newman black hole of mass $M$, charge $Q$ and angular momentum
$J=aM$ is described by
\begin{eqnarray}\label{kerr}
ds^2&=&-\left(1-\frac{2Mr-Q^2}{\rho^2}\right)dt^2-\frac{2a(2Mr-Q^2)\sin^2\theta}{\rho^2}dtd\phi\nonumber\\
&~&+\frac{\rho^2}{\Delta}dr^2+\rho^2d\theta^2+\frac{\sin^2\theta}{\rho^2}[(r^2+a^2)^2-a^2\Delta\sin^2\theta]d\phi^2,\nonumber
\end{eqnarray}
where
\begin{eqnarray}
\Delta&=&r^2-2Mr+a^2+Q^2,\nonumber\\
\rho^2&=&r^2+a^2\cos^2\theta.\nonumber
\end{eqnarray}
The location of the horizon, determined by $\Delta(r_{+})=0$, is
\begin{eqnarray}
r_{+}=M+\sqrt{M^2-Q^2-a^2}.
\end{eqnarray}
The horizon area $A$, surface gravity $\kappa$,  electric potential
$\phi$, and  angular velocity $\Omega$ are respectively given by
\begin{eqnarray}
A&=&4\pi(r_{+}^2+a^2),\nonumber\\
\kappa&=&\frac{r_{+}-M}{r_{+}^2+a^2}=\frac{4\pi(r_{+}-M)}{A},\nonumber\\
\phi&=&\frac{r_{+}Q}{r_{+}^2+a^2}=\frac{4\pi r_{+}Q}{A},\nonumber\\
\Omega&=&\frac{a}{r_{+}^2+a^2}=\frac{4\pi a}{A}.
\end{eqnarray}
These quantities yield the following relation\cite{beken}
\begin{eqnarray}\label{firstlaw3}
dM=\frac{\kappa}{8\pi}dA+\phi dQ+\Omega dJ.
\end{eqnarray}
It is the first law of a Kerr-Newman black hole, in the context of
mechanics.

 In the previous
subsection, we suggest that for a particle captured by black hole,
the position uncertainty $\Delta x$ yield
\begin{eqnarray}\label{rhox}
\Delta x\leq 2\rho_0,
\end{eqnarray}
where $\rho_0$ is a scale relevant  to the black hole. For a static
and spherically symmetric black hole, this characteristic size is
identified with the twice radius of the horizon. We are confronted
with a question of understanding the meaning of $\rho_0$, when a
rotating black hole is considered. At a first glimpse,  it appears
natural that  $\rho_0$ is represented by $r_{+}$. However, this
proposal is doubtable, although it is workable for the static and
spherically symmetric cases. This is because  the spatial part of
Boyer-Lindquist coordinates are different from ordinary polar
coordinates. For instance, in a rectangular coordinates $(X,Y,Z)$,
$r=const$ represents an ellipsoid rather than a sphere. Concretely
speaking, the coordinates $(r,\theta,\phi)$ are related to the
rectangular coordinates by\cite{boyer,ruffini,liang}
\begin{eqnarray}\label{boyer1}
X&=&\sqrt{r^2+a^2}\sin\theta\cos\phi^{*},\nonumber\\
Y&=&\sqrt{r^2+a^2}\sin\theta\sin\phi^{*},\\
 Z&=&r\cos\theta,\nonumber
\end{eqnarray}
where
\begin{eqnarray}
\phi^{*}=\phi-\tan^{-1}\frac{a}{r}-a\int_{\infty}^{r}\frac{dr}{\Delta}.\nonumber
\end{eqnarray}
Following from (\ref{boyer1}), we obtain
\begin{eqnarray}\label{ellips}
\frac{X^2+Y^2}{r^2+a^2}+\frac{Z^2}{r^2}=1,
\end{eqnarray}
which is axially symmetric. Obviously,  the surface of a Kerr-Newman
black hole($r\rightarrow r_{+}$) is a confocal ellipsoid. This
ellipsoid is characterized by  two scales: $r_{+}$ and
$\sqrt{r_{+}^2+a^2}$. Which is the characteristic size that
represents $\rho_0$? In order to minimize $\Delta A$, we choose the
latter, i.e.
\begin{eqnarray}\label{rho0}
\rho_0=\sqrt{r_{+}^2+a^2}.
\end{eqnarray}
One of the evidences for (\ref{rho0}) is that the absorption cross
section for a Kerr-Newman black hole  is proportional to its
area\cite{strominger}, $\sigma_{abs}\sim A=4\pi\rho_0^2$, which can
be interpreted by the aid of a two-body process in an effective
string theory that describes the collective excitations of the black
hole at weak coupling\cite{maldacena, das, cvetic}. This means that
$\rho_0$ is indeed a characteristic size in the absorption process.

 Furthermore, as argued immediately, (\ref{rho0}) is a reasonable choice in the sense of thermodynamics,
 which can be explained along another line of arguments. Let us return to (\ref{rhox}),
 where $\rho_0$ is to be determined. Replacing $r_0$ with $\rho_0$ and redoing the procedure from
(\ref{restrict1}) to (\ref{dads}), we obtain
\begin{eqnarray}\label{newdads}
\frac{(\Delta A)_{min}}{(\Delta S)_{min}}
=\frac{2\gamma_1\hbar}{\alpha\ln
2}(\rho_{0}^2-\rho_{0}\sqrt{\rho_{0}^2-\alpha}).
\end{eqnarray}
If $\rho_0$ is identified directly with $r_{+}$, (\ref{newdads})
becomes
\begin{eqnarray}
\frac{(\Delta A)_{min}}{(\Delta S)_{min}}
&=&\frac{2\gamma_1\hbar}{\alpha\ln 2}(r_{+}^2-r_{+}\sqrt{r_{+}^2-\alpha})\nonumber\\
&=&\frac{\gamma_1\hbar}{2\pi\alpha\ln 2}\left[A-4\pi
a^2-\sqrt{(A-4\pi a^2-2\alpha\pi)^2-4\alpha^2\pi^2}\right],\nonumber
\end{eqnarray}
which means that the entropy depends on two quantities: $A$ and $a$.
This contradicts Bekenstein's assumption that the entropy of a black
hole is a function only of its area\cite{beken}. This  would lead to
a deduction incompatible with thermodynamics. Supposing $S=S(A, a)$,
we have
\begin{eqnarray}\label{ds}
dS&=&\frac{\partial S}{\partial A}dA+\frac{\partial S}{\partial
a}da\nonumber\\
&=&\frac{\partial S}{\partial A}dA+M^{-1}\frac{\partial S}{\partial
a}(dJ-adM).
\end{eqnarray}
 In a reversible process, the black hole area is unchanged\cite{christ1,christ2},
 $dA=0$. So the change in black hole mass is attributed  to the work done by an external agent
 which changes the black hole's charge and angular momentum, and the first law (\ref{firstlaw3}) becomes
\begin{eqnarray}
dM=\phi dQ+\Omega dJ.\nonumber
\end{eqnarray}
Eq.(\ref{ds}) is therefore rewritten as
 \begin{eqnarray}
 dS=M^{-1}\frac{\partial S}{\partial
a}\left[(1-a\Omega)dJ-a\phi dQ\right],
 \end{eqnarray}
  which means $dS\neq 0$ if $(\partial S/\partial a)_A\neq 0$, since $Q$ and $J$ are independent variables.
  Especially for a neutral black hole, we have
\begin{eqnarray}
 dS=M^{-1}\frac{r_{+}^2}{r_{+}^2+a^2}\frac{\partial S}{\partial
a}dJ.\nonumber
 \end{eqnarray}
The black hole entropy could be increased by an external agent which
increases the angular momentum reversibly, if $(\partial S/\partial
a)_A\neq 0$. This means that the entropy is not invariant in such a
reversible process. This contradicts with the basic concept of
thermodynamics. The crucial reason is that $\rho_0$ is improperly
interpreted as $r_{+}$.

 What is $\rho_0$? Enlightened by the above discussion, $\rho_0$ should be  unchanged in a reversible process.
 This is required by the fact that $S$ and $A$ are invariant
 in the same process. Following from (\ref{newdads}),
  the black hole entropy is expressed as $S=S(A, \rho_0)$, so we
  have
\begin{eqnarray}
dS&=&\frac{\partial S}{\partial A}dA+\frac{\partial S}{\partial
\rho_0}d\rho_0.\nonumber
\end{eqnarray}
$d\rho_0=0$ as $dS=dA=0$, namely, $\rho_0$ is an invariant in a
reversible process. For a rotating black hole in a reversible
process, its irreducible mass $M_{ir}$ is unchanged
\cite{christ1,christ2}, where
\begin{eqnarray}\label{mir}
M_{ir}=\sqrt{\frac{A}{16\pi}}=\frac{1}{2}\sqrt{r_{+}^2+a^2}.
\end{eqnarray}
This similarity implies that $\rho_0$ should be interpreted as the
black hole irreducible mass, $\rho_0\sim M_{ir}$. This can be
understood  in another manner. We notice that (\ref{newdads})
involves three quantities of a black hole: the area $A$, entropy $S$
and the characteristic size $\rho_0$. $\rho_0$ is thus related not
only
 to $A$ but also to $S$. In other words, $\rho_0$ is a bridge
which crosses the gap between $A$ and $S$, hence it has  geometric
and thermodynamic meanings. The black hole irreducible mass agrees
with this requirement. On one hand, $M_{ir}$ is related to the area
by (\ref{mir}). On the other hand, $M_{ir}$ is the energy that can
not be extracted by a classical process( e.g. Penrose process). In
the thermodynamic sense, $M_{ir}$ corresponds to the degraded energy
that can not be transformed into work\cite{beken}. As a measure for
the degradation of energy, the entropy is related to the irreducible
mass by $S=S(M_{ir})$.  Thus $\rho_0$ is  endued with geometric and
thermodynamic meanings by identifying it with $M_{ir}$. Therefore,
(\ref{rho0}) is a natural choice in the context of thermodynamics.

Replacing $r_0$ with $\rho_0$ and doing the discussion parallel to
the previous subsection, we obtain the temperature, entropy and heat
capacity of a Kerr-Newman black hole, i.e.
\begin{eqnarray}\label{temp3}
T&=&\frac{\hbar^{\prime}\kappa}{2\pi}, \\
S&=&\frac{\pi}{2\hbar}\left[\rho_0^2+\rho_0\sqrt{\rho_0^2-\alpha}-\alpha\ln(\rho_0+\sqrt{\rho_0^2-\alpha}
)\right],\nonumber\\
C&=&C_0f\left[\frac{\hbar^{\prime}}{\hbar}f-C_0\Delta\hbar\right]^{-1},\nonumber
\end{eqnarray}
where $\rho_0=\sqrt{r_{+}^2+a^2}$, and
\begin{eqnarray}
\hbar^{\prime}&=&\frac{2\hbar}{\alpha}\left(\rho_0^2-\rho_0\sqrt{\rho_0^2-\alpha}\right),\nonumber\\
f&=&\pi \rho_0\sqrt{\rho_0^2-\alpha}.\nonumber
\end{eqnarray}
Obviously, the expressions for these thermodynamic quantities are
similar to the static and spherical black holes.

\section{Black hole remnant}
GUP provides a possible mechanism to prevent a black hole from
complete evaporation\cite{adler2}. Let us talk about our
understanding of this suggestion. In the static and spherically
symmetric situation, the black hole radius is restricted by GUP and
must be greater than a minimum value($r_0\geq\sqrt{\alpha}$). This
restriction is necessary
 to (\ref{temp2}) and (\ref{entropy}),  otherwise the two expressions
 would lose their physical meanings. According to (\ref{heat}), the black
hole's  heat capacity vanishes when its radius approaches the
minimum value. The zero heat capacity means that no particle is
emitted from the horizon, which suggests  a black hole remnant. As a
comparison, we temporarily ignore the GUP's impacts and first
consider a Schwarzschild black hole in the standard framework. Its
heat capacity is $C_0\propto -M^2$. Due to the negative heat
capacity, the black hole will evaporate to zero mass, and then
$C_0\rightarrow 0$.  However, GUP makes the situation different: (i)
the black hole acquires a nonzero minimum size of $\sqrt{\alpha}$;
(ii) the heat capacity vanishes as $r_0\rightarrow\sqrt{\alpha}$. In
comparison with the standard result that $C_0\rightarrow 0$ as
$M\rightarrow 0$, we find that GUP elevates the black hole's ``zero
point energy" to a new scale determined by $r_0=\sqrt{\alpha}$.

 However, Hawking radiation  makes a black hole unstable. A realistic black hole is thus
 non-stationary, and should be approximately described by, for example,  Vaidya metric\cite{vaidya}
\begin{eqnarray}
ds^2=-\left[1-\frac{2m(v)}{r}\right]dv^2+2dvdr+d\Omega^2,\nonumber
\end{eqnarray}
where $v$ is the Eddington-Finkelstein coordinate which denotes the
retarded time. Its horizon is located by\cite{hiscock}-\cite{zdai}
\begin{eqnarray}
r_H=\frac{2m}{1-4\dot{m}},\nonumber
\end{eqnarray}
where $\dot{m}=dm/dv$ is the mass loss rate. GUP restricts the black
hole's radius by $r_H\geq \sqrt{\alpha}$, i.e.
\begin{eqnarray}\label{2m}
\frac{2m}{1-4\dot{m}}\geq \sqrt{\alpha}.
\end{eqnarray}
For an evaporating black hole, its mass always decreases with time,
i.e. $\dot{m}< 0$.  Considering (\ref{2m}), we obtain
\begin{eqnarray}
0\leq -4\dot{m}\leq\frac{2m(v)}{\sqrt{\alpha}}-1,\nonumber
\end{eqnarray}
where $\dot{m}=0$ denotes  a black hole which stops evaporating.
Obviously, $\dot{m}\rightarrow 0$ as
$m(v)\rightarrow\sqrt{\alpha}/2$. Hawking radiation is  shut off
when the black hole evaporates to a Planck scale mass.

 Black hole remnant has been suggested as an information
loss reposition to resolve the black hole information
problem\cite{ahar, preskill}. The remnant is assumed to retain the
large information of the initial black hole although it has a small
size and a tiny mass. However, this idea is questionable since it
violates Bekenstein's entropy bound\cite{beken2}, $S\leq 2\pi
E\ell/\hbar$, where $E$ denotes the energy of the system of interest
and  $\ell$  the size. Following from this bound, the remnant's
information content is  a few bits at most. It is too tiny to
resolve the  information loss problem.

Can the situation be improved when a weaker constraint is
considered? In an asymptotical de Sitter spacetime, the entropy of a
matter system is restricted by the so-called D bound\cite{bousso}
\begin{eqnarray}\label{dbound1}
S_m\leq \frac{1}{4}(A_0-A_c),
\end{eqnarray}
which is derived from the generalized second law via a Geroch
process, where $A_0$ and $A_c$ are the areas of the cosmological
horizons of
 pure and asymptotical de Sitter spacetimes respectively. This consideration is motivated by the astronomical
 observation that the
current universe is dominated by the dark energy. Cosmological
constant, $\Lambda$, is the simplest candidate for the dark energy.
The information capacity of black hole remnant deserves to be
seriously considered in the de Sitter spacetime. D-bound takes the
form\cite{bousso}
\begin{eqnarray}\label{dbound2}
S_m\leq \pi r_gr_c,
\end{eqnarray}
when the gravitational radius  of the matter system($r_g$) is much
less than the radius  of the cosmological horizon($r_c$). For a
black hole remnant, its gravitational radius acquires  the minimum
value determined by GUP. Replacing $\sqrt{\alpha}$ for $r_g$,
(\ref{dbound2}) is deduced to
\begin{eqnarray}\label{dbound3}
S_{r}\leq\sqrt{\alpha}\pi r_c<\pi \sqrt{\frac{3\alpha}{\Lambda}},
\end{eqnarray}
 where we have considered
$r_c<r_0=\sqrt{3/\Lambda}$.  Following from quantum statistical
mechanics, the entropy bound (\ref{dbound3}) means that the number
of the internal states of a black hole remnant is less than
$\exp(\pi\sqrt{3\alpha/\Lambda})$. In other words, the information
capacity of a black hole remnant in the de Sitter spacetime is
restricted by the bound (\ref{dbound3}), which is concretely
determined by  the cosmological constant. In Planck units, the
observed value of $\Lambda$ is about $\sim 10^{-120}$, and then
$S_r$ acquires the value of $10^{60}$ bits at most. Comparing with
Bekenstein's bound,  D-bound increases the remnant's information
capacity dramatically. However, the situation is  still not
optimistic. Considering a black hole of initial mass $M_0$, its
entropy is $S_0\approx 4\pi M_0^2$, which measures the total
information hidden at the moment of collapse. For a solar mass black
hole, its entropy is about $10^{76}$ bits, which is about 16 orders
greater than $S_r$. This means that the remnant cannot retain the
total information content of the initial black hole. The discrepancy
becomes more serious when the larger black holes are considered. In
order for the entropy bound (\ref{dbound3}) to be workable, the
black hole mass must yield
\begin{eqnarray}
M_0<\left(\frac{3\alpha}{16\Lambda}\right)^{1/4}\sim 10^{30}m_p\sim
10^{25}g,\nonumber
\end{eqnarray}
which is 8 orders less than the solar mass. Obviously, this mass
scale rules out the most black holes in the universe. We therefore
arrive at a conclusion that black hole remnants might not serve to
resolve the information paradox.

\section{Summary}
This research explores an alternative expression for black hole
thermodynamics in the sense of GUP (\ref{gup}). We try to extend the
argument of Ref.\cite{adler2} to more general black holes. We first
consider a static and spherically symmetric black hole, and work out
the temperature, entropy and heat capacity of it. These quantities
are expressed by (\ref{temp2}),(\ref{entropy}) and (\ref{heat}). The
similar expressions are also valid to a Kerr-Newman black hole, when
$r_0$ is replaced by $\rho_0$. For example, the temperatures of both
black holes, (\ref{temp2}) and (\ref{temp3}), can be expressed as a
unified form, $T=\hbar^{\prime}\kappa/2\pi$, if $r_0$ and $\rho_0$
are written as $\sqrt{A/4\pi}$.

Our argument is based on the analysis of a gedanken experiment that
a particle is absorbed by black hole. Different from the existing
literature, we suggest that the black hole irreducible mass
represent the characteristic size in the absorption process, which
restricts the position uncertainty of the particle falling in the
black hole. This suggestion follows from two evidences: (a) the
absorption cross section of a black hole is proportional to its
area; (b) the entropy is an invariant in a reversible process. This
suggestion can be applied to a rotating black hole and is compatible
with the static and spherically symmetric black hole.

GUP (\ref{gup}) does not depend on the concrete form of a black
hole. This implies that the temperature expression (\ref{temp3}) may
be applied to more black holes, such as the black holes with
dilaton. Following from the previous discussion, this generalization
should be based on two assumptions:(i) the first law
(\ref{firstlaw1}) and the increase in area (\ref{minia0}) are still
workable; (ii) there exists a characteristic size in the absorption
process, which minimizes $\Delta A$ and is an invariant in a
reversible process.

 Following from D-bound, a remnant cannot retain the total
information of the initial black hole. Is it possible for the
D-bound to be corrected by GUP? Since there are some similarities(in
the sense of thermodynamics) between the cosmological horizon and
black hole, we guess that a log type correction, $\ln (A_c/A_0)$,
might appear in the r.h.s of (\ref{dbound1}). However,
 this correction is too tiny to overset the conclusion  from the D-bound.

\section*{Acknowledgments}
The authors would like to thank the anonymous referee who stimulates
an essential improvement in this work.  The authors also thank Prof.
Yi Ling for a part of this work. This research is supported by NSF
of China(Grant Nos.10673001, 10875057), NSF of Jiangxi
province(Grant No.0612038), the key project of Chinese Ministry of
Education(No.208072) and Fok Ying Tung Education
Foundation(No.111008). We also acknowledge the support by the
Program for Innovative Research Team of Nanchang University.

\end{document}